# Remote patient monitoring using artificial intelligence: Current state, applications, and challenges

Thanveer Shaik[1] | Xiaohui Tao[1] | Niall Higgins[2,3] | Lin Li[4] | Raj Gururajan[5] | Xujuan Zhou[5] | U. Rajendra Acharya[6]

[1]School of Mathematics, Physics and Computing, University of Southern Queensland, Toowoomba, Australia

[2]Metro North Hospital and Health Service, Royal Brisbane and Women's Hospital, Brisbane, Australia

[3]School of Nursing, Queensland University of Technology, Brisbane, Australia

[4]School of Computer and Artificial Intelligence, Wuhan University of Technology, Wuhan, China

[5]School of Business, University of Southern Queensland, Springfield, Australia

[6]School of Science and Technology, Singapore University of Social Sciences, Singapore

**Correspondence**
Thanveer Shaik, School of Mathematics, Physics and Computing, University of Southern Queensland, Toowoomba, Queensland, Australia.
Email: thanveer.shaik@usq.edu.au

**Edited by:** Tianrui Li, Associate Editor and Witold Pedrycz, Editor in Chief

**Abstract**

The adoption of artificial intelligence (AI) in healthcare is growing rapidly. Remote patient monitoring (RPM) is one of the common healthcare applications that assist doctors to monitor patients with chronic or acute illness at remote locations, elderly people in-home care, and even hospitalized patients. The reliability of manual patient monitoring systems depends on staff time management which is dependent on their workload. Conventional patient monitoring involves invasive approaches which require skin contact to monitor health status. This study aims to do a comprehensive review of RPM systems including adopted advanced technologies, AI impact on RPM, challenges and trends in AI-enabled RPM. This review explores the benefits and challenges of patient-centric RPM architectures enabled with Internet of Things wearable devices and sensors using the cloud, fog, edge, and blockchain technologies. The role of AI in RPM ranges from physical activity classification to chronic disease monitoring and vital signs monitoring in emergency settings. This review results show that AI-enabled RPM architectures have transformed healthcare monitoring applications because of their ability to detect early deterioration in patients' health, personalize individual patient health parameter monitoring using federated learning, and learn human behavior patterns using techniques such as reinforcement learning. This review discusses the challenges and trends to adopt AI to RPM systems and implementation issues. The future directions of AI in RPM applications are analyzed based on the challenges and trends.

This article is categorized under:
  Application Areas > Health Care
  Technologies > Artificial Intelligence
  Technologies > Internet of Things

**KEYWORDS**
artificial intelligence, IoT, noninvasive technology, remote patient monitoring

# 1 | INTRODUCTION

Remote patient monitoring (RPM) is a rapidly growing field in healthcare that is designed to assist clinicians with additional support to provide care in a range of general hospital medical and surgical wards and using flexible materials for wearable sensors (Joshi et al., 2021; Liu, Wang, et al., 2022; Weenk et al., 2020). This is achieved by incorporating new Internet of Things (IoT) methodologies in healthcare such as telehealth applications (Heijmans et al., 2019), wearable devices (Dias & Cunha, 2018), and contact-based sensors (Malasinghe et al., 2017). RPM is commonly used to measure vital signs or other physiological parameters such as motion recognition that can assist with clinical judgments or treatment plans for conditions such as movement disorders or psychological conditions (Shaik, Tao, Higgins, Gururajan, et al., 2022; Shaik, Tao, Higgins, Xie, et al., 2022).

Artificial intelligence (AI) algorithms have been employed to perform analysis of medical images and correlate symptoms and biomarkers from clinical data to characterize an illness and its prognosis (Miller & Brown, 2018; Schnyer et al., 2017). There is immense potential for AI to benefit healthcare service delivery and clinicians are exploring a variety of practical issues for assessing the risk of disease, ongoing patient care, and how AI can help clinicians to alleviate or reduce complications in illness progression (Torous et al., 2018). Medical research is also benefitting from AI by helping to expedite genome sequencing and the development of new drugs and treatments from the knowledge that previously was not possible to obtain or observe from such complex data. Machine learning, a subset of AI, can potentially assist clinicians in interpreting complex data in a relatively short period using specialized algorithms (Helm et al., 2020; Krittanawong et al., 2022). They can assist with a patient assessment to help predict early deterioration of their health status and even classify their types of motion or activities (Z. Liu, Zhu, et al., 2022; Huang et al., 2022). These AI algorithms can process large datasets to recognize and learn complex patterns for decision-making (Dean et al., 2022). Recent increases in computational speed have led to the development of even more powerful artificial neural networks and deep learning algorithms that can handle and optimize very complex datasets (Bini, 2018; Kalfa et al., 2020). Many routine tasks can be automated by incorporating an IoT model with a centralized control unit and interface. This could potentially avoid human errors, and increase patient safety (Tandel et al., 2022).

RPM has traditionally been applied to monitoring patients in rural areas remotely using telehealth technology, monitoring chronically ill people, and the elderly at home using wearable devices or sensors, but the nonintrusive aspects are also attractive for use in hospitals for post-surgery patients, and those in intensive care units using wireless body sensors. It is possible to enhance these monitoring systems to the next level by introducing noninvasive digital technologies which permit patients' daily activities. To support healthcare professionals in visualizing the health status of patients based on vital signs and activity recognition, machine learning (ML) and AI can be implemented as shown in Figure 1. These types of applications can present data related to diagnosing and predicting patient health status and assist with clinical decision-making. This review is motivated by potential advancements in healthcare using AI and machine learning to transform existing traditional medical practices.

This review aims to investigate technologies adopted in current RPM systems for noninvasive techniques. Current trends in RPM and applications of AI to monitor vital signs, physical activities, emergency events, and chronic diseases of patients and assist clinicians to diagnose and provide efficient care. The impact of AI on RPM applications for early detection of health deterioration, personalized monitoring, and adaptive learning are discussed. Finally, the current challenges to the widespread adoption of remote monitoring with AI or machine learning in healthcare are presented and identify what is being done to address these. The contributions of this study are:

- AI impact of RPM applications is investigated and stressed the need for early detection of health deterioration.
- Traditional machine learning and deep learning applications in RPM are investigated.
- Comprehensive review of advanced technologies such as video-based monitoring, IoT-enabled devices, cloud, edge, fog, and blockchain and AI methodologies such as reinforcement learning, and federated learning adopted by RPM systems.
- Challenges in adopting AI-enabled RPM are investigated, and their trends are explored.

The review is organized as follows: Section 2 presents the research explored in this study, search strategies, and inclusion criteria. Section 3 presents advanced RPM architectures including telehealth, IoT, cloud, fog, edge, and blockchain technologies. The scope of AI in RPM applications like monitoring vital signs, physical activities, emergencies, and chronic diseases are discussed in Section 3. In Section 4, the impact of AI on RPM has been discussed. Section 5 describes the challenges involved in adopting AI or machine learning to monitor patients. Section 6 concludes the article with a finding summary and future work with recommendations.

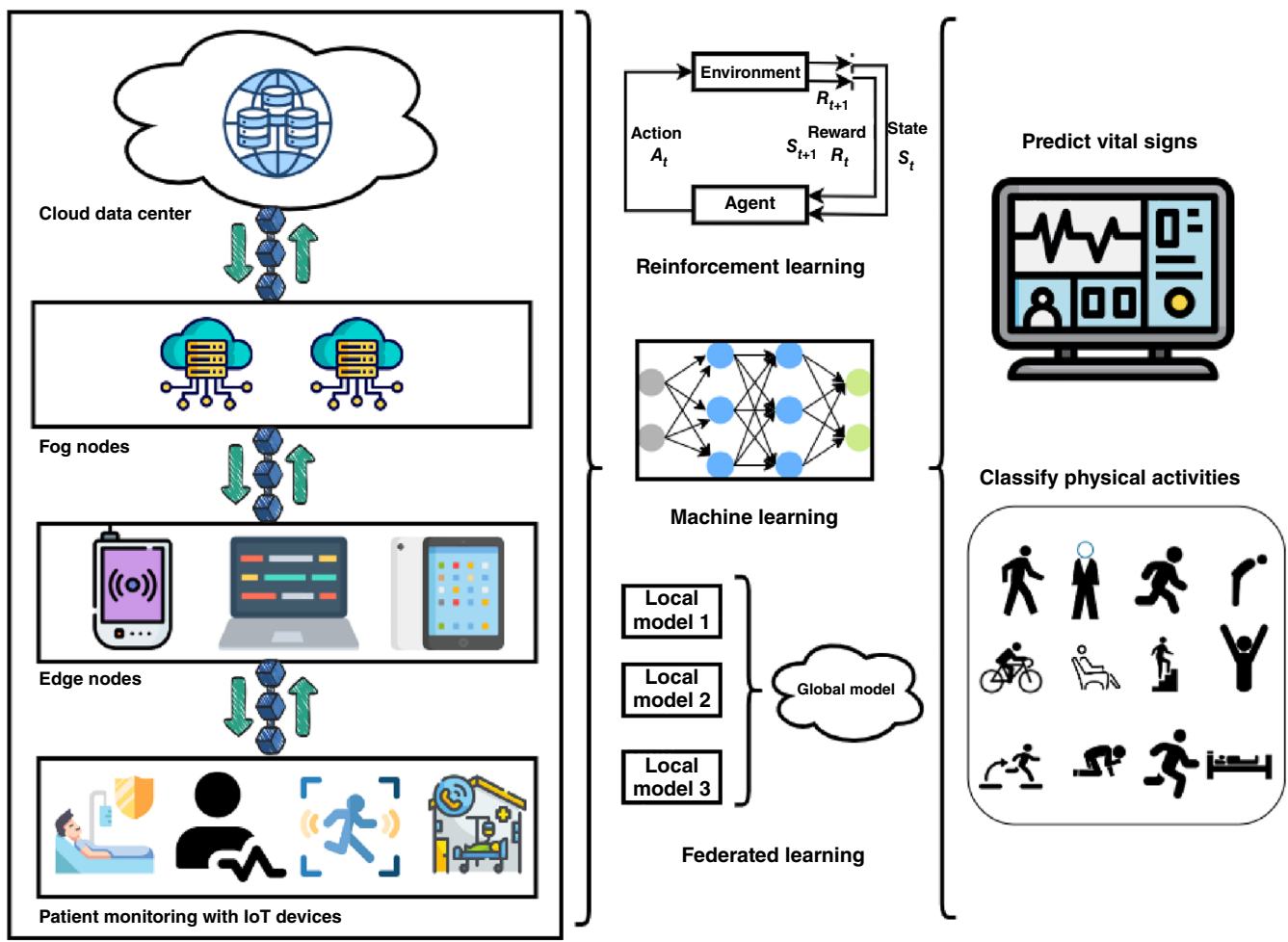

**FIGURE 1** Artificial intelligence-enabled remote patient monitoring architectures

## 2 | SEARCH STRATEGY AND SELECTION CRITERIA

The objective of this review is to identify journal articles, review articles, and conference papers related to the role of AI in monitoring a patient's health status. This can be done using IoT devices in geographically remote settings or more locally through nontouch techniques. In doing so, the review will seek to address the following research questions:

**RQ1.** What technologies have transformed conventional manual patient monitoring in hospitals?

**RQ2.** How has AI transformed the RPM with its advancements and their impacts?

**RQ3.** What are the challenges in adopting AI for RPM systems and learning healthcare data?

**RQ4.** What are the existing trends in RPM systems for using AI?

**TABLE 1** Selected limits for database searches

| Inclusion criteria | Exclusion criteria |
| --- | --- |
| Journal Article | Books |
| Review Article | Book chapters |
| Conference Paper | Abstracts |
| Conference Paper Review | Short survey Editorial |
| Published between 2016 and 2022 | Letter |
| Literature in English | Research works related to infants, neonates |
| Outpatients and inpatients | Experiments on animals |
| Employs AI & ML | Research work without AI & ML |
| Experiments on adult and elderly patients | Image processing techniques |

Abbreviations: AI, artificial intelligence; ML, machine learning.

## 2.1 | Information sources

Literature was retrieved from the following bibliographic databases: Web of Science, Scopus, Springer, ACM Digital Library, IEEE Xplore, Pub-Med, Science Direct, and Multidisciplinary Digital Publishing Institute (MDPI). Search strategies were defined using keywords, Boolean operators, truncation, and wildcards. Each database was filtered to search the keywords and their combinations in the title, abstract, and keywords. Results were sorted by relevance, and the first 10 results were checked to ensure that a combination of search terms retrieved articles relevant to the research questions. Finally, the results were exported to EndNote and grouped for each database. Furthermore, the EndNote citations were exported to software called Rayyan (Ouzzani, Hammady, Fedorowicz, & Elmagarmid, 2016) to facilitate the screening and selection process. As databases could host articles published elsewhere, duplicate articles were excluded.

## 2.2 | Search strategy

Before defining the keywords, a random search was conducted to identify keywords that have been previously used to retrieve relevant articles on these topics. Once this was completed, the main concepts in each article title were categorized into five areas, and synonyms were created using a thesaurus. Table I presents the keywords used for each concept's search, with truncation and wildcards.

Boolean operators AND, and OR were used to form different combinations of keywords within the five key concepts. The final search string used in all the databases was:

> (patient? OR victim? OR case? OR subject? OR human?) AND (observ* OR monitor* OR audit* OR detect* OR estimat* OR forecast* OR check*) AND (remote OR distan* OR isolated OR inaccessib* OR outlying) AND (RFID OR sensor* OR wire* OR accelerometer OR doppler OR ECG OR radio* OR polysomno*) AND ("artificial intelligence" OR AI OR "machine learning" OR "neural networks")

## 2.3 | Selection criteria

As the review focused on the implementation of AI or machine learning in collaboration with information systems infrastructures, this study excluded research articles with continuous monitoring without AI or machine learning. Table 1 presents the chosen limits used to retrieve articles published between 2016 and 2021 and Figure 2 presents a PRISMA flowchart for the review process.

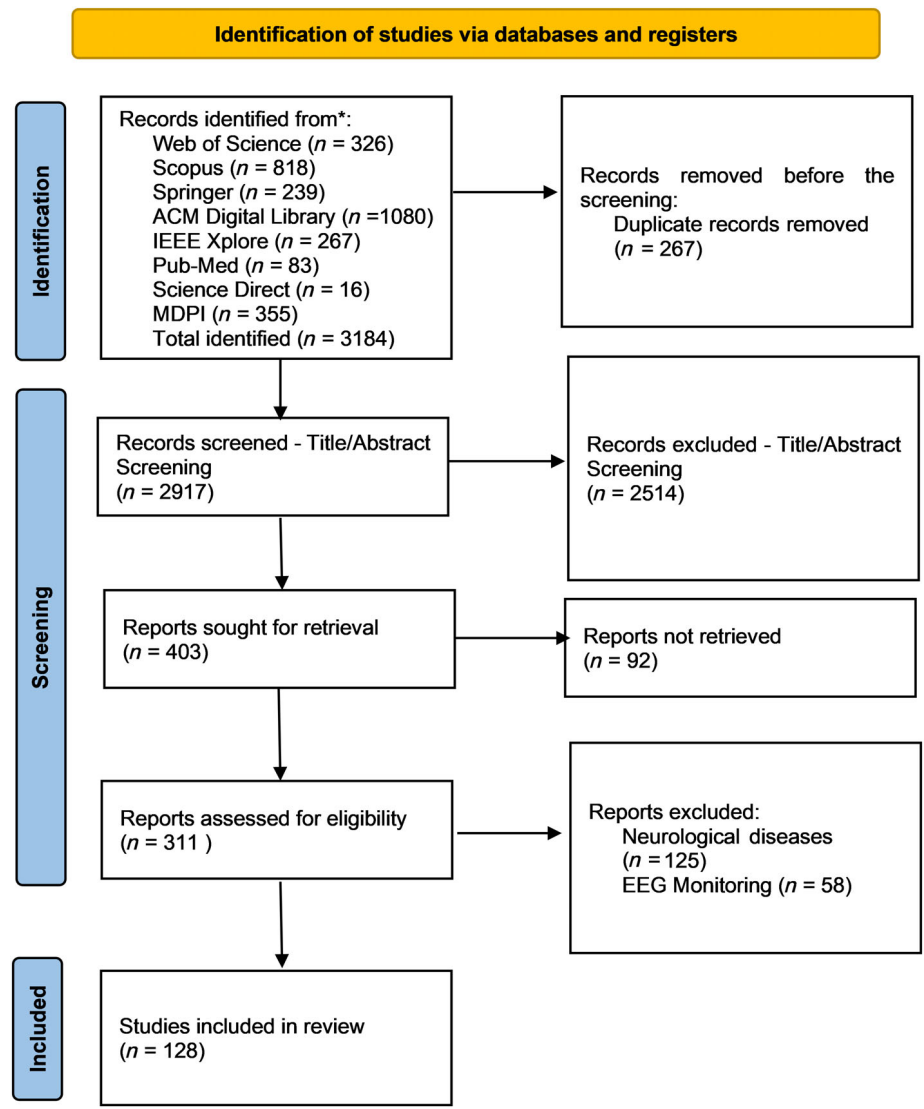

**FIGURE 2** Study flow diagram PRISMA-ScR (Page et al., 2021)

## 3 | REMOTE PATIENT MONITORING ARCHITECTURES

In hospitals, medical staff monitored a patient's health status regularly and manually maintained the records. Collecting patients' vital signs manually in hospitals depends on factors like clinical workload, staff working hours, patients' diagnosis, clinical leadership, and national guidance (Smith et al., 2017) and was limited due to the lack of resources. The patient monitoring was with invasive devices requiring patients' skin contact to estimate their vital signs. Technological advancements in data transmission have disrupted the healthcare industry with noninvasive devices without touching patients' bodies and provided opportunities to monitor patients continuously. The innovations have transformed the traditional patients' health status monitoring patterns and enabled to monitor of patients remotely in hospitals, patient care hospitals, age care facilities, and even in their homes. In this section, technology-enabled RPM architectures are discussed.

### 3.1 | Video-based monitoring

Telehealth monitoring allows patients to contact their doctors or medical staff via audio call or video call using smart devices. Snoswell et al. conducted a systematic review to measure the clinical effectiveness of telehealth applications. In their study, Snoswell et al. (2021) reviewed 38 meta-analysis articles published between 2010 and 2019 and covered 10 medical disciplines including multidisciplinary care and other specialized disciplines like cardiovascular disease and pulmonary diseases. The authors reported that the usage of telehealth has exponentially increased over the last decade and demonstrated examples from mental health support, pain management, blood pressure and glucose control, stroke management, and diagnostic services like dermatological and ophthalmic conditions.

As a result of the COVID-19 pandemic, telehealth became a common strategy for maintaining patients' and clinicians' safety. Machine learning and image processing techniques played a vital role in telehealth monitoring. The AI methods are capable of monitoring patients' vital signs such as heart rate, respiratory rate, oxygen saturation ($SpO_2$), cough analysis, and blood pressure. Rohmetra surveyed AI-enabled telehealth monitoring of vital signs and compared these with traditional methods of monitoring vital signs (Rohmetra et al., 2021). The image and video processing techniques in ML helped identify a region of interest (ROI) on the patient such as facial landmarks and then focused on the selected ROI to estimate vital signs that included heart rate, and respiratory rate. Bousefsaf et al. (2019) monitored the patterns of patients' pulse rates in an ROI of a video frame based on the fluctuations of movement during breathing. Based on the breathing patterns detected in video monitoring, Cho et al. (2017) developed a deep learning model, convolutional neural networks (CNN), which recognizes people's psychological stress levels. Cough analysis was performed based on auscultation sounds by employing a pretrained 3D ResNet18 neural network model to classify the sounds into disease categories. The model achieved 94.57% accuracy, 100% sensitivity, and 94.11% specificity. Heart rate, blood volume pulse, and $SpO_2$ were measured based on remote Photoplethysmography (rPPG) detected in a video frame captured by a standard smartphone camera (Khalid et al., 2022). The change in blood volume pulse causes blood absorption during a heartbeat was measured by focusing on forehead ROI using the Viola-Jones algorithm. The PPG signal extracted from the video and the ground truth blood pressure from the algorithm was fed into a feed-forward neural network model. The model achieved 85% accuracy in extracting the blood pressure. Laurie et al. further demonstrated how an algorithm specifically designed to control exposure time during video capture improves the accuracy of rPPG (Laruie et al., 2021).

Studies that explored the advantages and disadvantages of telehealth are also presented in Table 2. Telehealth cut down travel time, clinic visits, and extended time off work (Nord, Rising, Band, Carr, & Hollander, 2019). However, there are challenges associated with the benefits of telehealth monitoring. Overutilization or misuse of telehealth services has increased healthcare costs to providers (Busso et al., 2022). Telehealth monitoring has widened the disparities between rural and urban populations due to the accessibility of the internet and technology (Drake et al., 2019). Patient data security is another challenge in telehealth monitoring, which jeopardizes patients' health information without an end-to-end encrypted communication service (Fang et al., 2020).

Telehealth patient monitoring techniques have the potential to diagnose patients' health status. AI-enabled telehealth monitoring would be the more enhanced approach to classifying or predicting patients' vital signs.

## 3.2 | IoT-enabled devices

An IoT based real-time remote patient monitoring system would help achieve continuous patient monitoring (Yew et al., 2020). The majority of IoT technology systems have been developed for use in a hospital setting or a private dwelling. However, there are examples where a single system could be readily applied to both. Figure 3 presents an example of typical architecture that could be used for patient monitoring. The architecture is breakdown into three sections (Pan et al., 2020). Section A illustrates the wearable devices connected to patients to collect vital signs such as heart rate, pulse rate, respiratory rate, breathing rate, body temperature, and so on. In Section B, the collection will be stored in cloud services (Neto et al., 2017; Shao et al., 2020; Shi et al., 2020) for further analysis using machine learning methodologies that could predict or classify the patient data. The process could then estimate any abnormal events in the near future based on known threshold values of the vital signs and update medical staff or healthcare professionals (Ankita et al., 2021; Bekiri et al., 2020; C. Liu et al., 2019; Lin et al., 2018; Devi & Kalaivani, 2019; Efat et al., 2020; Shao et al., 2020) in Section C of the architecture. IoT has the potential to interconnect wearable sensors and their reader-antennas with a patient body to the monitoring network. The types of wearable vital signs sensing technologies, their architectures, and specifications range from physiological measurements, including electrocardiogram, blood oxygen saturation, blood glucose, skin perspiration, and capnography, to motion evaluation and cardiac implantable devices (Dias & Cunha, 2018). The devices can also take the form of wearable t-shirts, chest straps, or adhesive patches. Medical staff or healthcare professionals would take appropriate actions to treat the patient and avoid abnormal events.

TABLE 2 Telehealth monitoring

| References | Algorithm | Technology | Advantages | Disadvantages |
| --- | --- | --- | --- | --- |
| Bousefsaf et al. (2019) | 3D-CNN | RGB camera | • Improved access and timeliness of care<br>• Emergency preparedness<br>• Cost-effectiveness<br>• Reduced doctor-patient supply–demand mismatch | • Overutilization or misuse of telehealth services increase healthcare costs to providers.<br>• Disparities between rural and urban populations. |
| Cho et al. (2017) | CNN | Thermal camera | | |
| Khalid et al. (2022) | DFT, CWT | Sensor, RGB camera | | |
| Laurie et al. (2021) | Exposure control | Sensor, RGB camera | | |

Abbreviation: CNN, convolutional neural network.

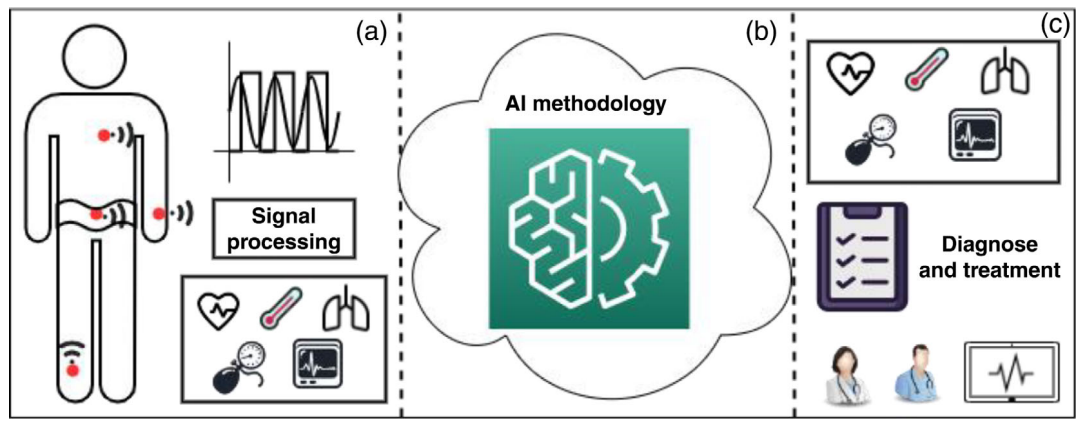

FIGURE 3 Patient monitoring architecture (inspired by Shao et al., 2020)

El-Rashidy et al. (2021) discussed trends and challenges of adopting a wireless body area network (WBAN), a subdomain of IoT which connects wireless sensors on a patient's body to the network. The WBAN challenges are transmission protocols, data privacy and security, interoperability, and integration. To transmit data from wireless sensors on a human body to local and global networks would need standard data transmission networks like ZigBee, Lora, Wi-Fi, and Bluetooth (Tripathi et al., 2020), and these networks have limitations in terms of energy and range of transmission. Data privacy and security challenges are inevitable in technology-enabled applications, and vast research on secured data transmission or data transaction processes and decentralized or distributed technology is being conducted to overcome the security and privacy challenges. Integration of sensors and remote devices led to sensor interoperability and combining heterogeneous data resources led to data interoperability in wireless sensor networks.

Sensors in RPMs require patients' skin contact to retrieve vital signs, limiting the daily activities of chronically diseased patients. Adopting radio frequency identification (RFID) technology overcomes the challenges of incisiveness in RPM. RFID technology has battery-powered active tags and battery-less passive tags that work on Near-field Coherent Sensing (NCS) principle. To enhance patient comfort and be less restrictive to their daily activities, noninvasive digital technology with NCS has been proposed and developed by Cornell University researchers (Hui & Kan, 2017; Sharma & Kan, 2018). NCS is a method developed by Hui and Kan (2017) that directly modulates the mechanical motion on the surface and inside a body onto multiplexed radio signals. This is integrated with a unique digital identification. In this mechanism, electromagnetic energy is directed into body tissue which reflects back-scattered signals from internal organs and is implicitly amplified. Small mechanical motions inside the body that have a shorter wavelength can be rendered into a large phase variation to improve sensitivity. NCS mechanisms were deployed to monitor vital signs, score sleep (Sharma & Kan, 2018), and accurately extract heartbeat intervals (Hui & Kan, 2018).

Passive RFID tags can be deployed into garments at the chest and wrist areas. This is where the two multiplexed far-field back-scattering waveforms are collected at the reader to retrieve the blood pressure, heart rate, and respiration rate. This could minimize deployment and maintenance costs. Hui and Kan (2017) found that to maximize reading range and immunity to multipath interference caused by indoor occupant motion, active tags could be placed in the

TABLE 3 Internet of Things (IoT) devices monitoring

| References | Algorithm | Technology | Advantages | Disadvantages |
| --- | --- | --- | --- | --- |
| El-Rashidy et al., 2021 | | Wireless sensors, ZigBee, Lora, Wi-Fi, Bluetooth | • Enable personalized health monitoring<br>• Noninvasive monitoring<br>• Continuous health monitoring | • Patient privacy concerns<br>• High dependence on the internet |
| Sharma & Kan, 2018 | SVM model | NCS, RFID Passive tags | | |
| Hui & Kan, 2018 | SVM model | NCS, RFID Passive tags | | |

Abbreviation: SVM, support vector machines.

front pocket. Also, placing the tags in the wrist cuff to measure the antenna reflection due to NCS. With this, vital signals can be sampled and transmitted digitally. Table 3 presents the IoT devices used by the research community for patients' monitoring, and their advantages and disadvantages are explained.

## 3.3 | Cloud computing

Cloud computing is an essential component of continuous patient monitoring systems. The huge amount of data generated for each patient from the IoT devices in RPMs needs storage to share data between different parties and analyze trends (Zamanifar, 2021). Cloud computing technology is a powerful platform that holds servers, databases, networking, software, and intelligence online (over the internet) for faster innovation and flexible resources. Iranpak et al. (2021) used the features of cloud computing features and developed a patient monitoring system based on IoT devices. The transmission of the data in the IoT platform to the cloud using the Fifth Generation Internet (5G) network. A deep learning neural network model, long short-term memory (LSTM), was used to monitor patients and classify their health conditions. The proposed deep learning model outperformed baseline models with an accuracy of 97.13%. Cloud computing uses a centralized data server to manage large amounts of data from all IoT devices. Integrating the IoT platform with cloud computing raises concerns about latency, real-time response delays, bandwidth overuse, and data security. This led to decentralized or distributed computing approaches like fog computing and edge computing, in which cloud services are brought close to IoT networks and overcome the cloud computing challenges (Pareek et al., 2021). The major concern of data security in cloud computing can be addressed by encrypting the data at the IoT device level and then sending data to cloud storage for data analysis. Siam et al. (2021) proposed the advanced encryption standard (AES) algorithm to encrypt the vital signs retrieved from a patient and send the encrypted data to the cloud. This allows only trusted medical organization servers to access the data with the appropriate decryption key, which will be kept secret between the system and the healthcare center. The proposed approach outperforms the commercial devices available on the market with minimal root-mean-square error, mean absolute error, and mean relative error of 0.012, 0.009, and 0.003, respectively.

## 3.4 | Fog and edge computing

Fog computing is an extension of cloud computing, which takes cloud computing services closer to IoT devices. Advancements in applications of IoT, and integrated cloud computing such as real-time monitoring of patient vital signs, physical activities have increased threats like security, performance, latency, and network breakdown to cloud computing (Sabireen & Venkataraman, 2021). Fog computing is a distributed or decentralized virtual network to act as a medium between IoT devices and the cloud (Alwakeel, 2021). Pareek et al. (2021) discussed IoT-Fog-based system architectures in healthcare. The delay in real-time responses and latency issues in cloud computing can be addressed by deploying fog nodes that analyze the data from the IoT platform with minimum delay time (Q. Qi & Tao, 2019). The fog computing architecture provides real-time analysis and security of data by preserving sensitive data and performing calculations closer to the IoT platform.

Similarly, cloud computing services are further pushed closer to the edge of the networks or IoT devices by introducing another decentralized or distributed concept called edge computing. The edge computing operations are executed in intelligent devices like programmable controllers, which read IoT devices (Alwakeel, 2021). Edge computing nodes

TABLE 4 Cloud/Fog/Edge monitoring

| References | Algorithm | Technology | Advantages | Disadvantages |
| --- | --- | --- | --- | --- |
| Uddin (2019) | LSTM model | IoT devices, edge computing | • Real-time automated analytics | • Privacy and security of patient data |
| Vimal et al. (2021) | CNN | IoT devices, Edge/Fog/Cloud computing | • Cloud computing services closer to patients | • Domain knowledge training for medical staff |
| Siam et al. (2021) | AES algorithm | IoT devices, cloud computing | • Decentralized network support personalized monitoring | |

Abbreviations: AES, advanced encryption standard; CNN, convolutional neural network; LSTM, long short-term memory.

deploy intermediate nodes closer to the network with storage and computation capabilities. The cloud/fog/edge monitoring architectures enable real-time monitoring with a decentralized approach for personalized care, but the technologies have their disadvantages, as shown in Table 4.

Uddin (2019) proposed a wearable sensor-based system with an AI-enabled edge device for patients' physical activity prediction. The graphics processing unit in the edge device was used for faster computation results. A deep learning LSTM model was used in the edge for the physical activity classification. The model achieved an accuracy of 99.69% mean prediction performance compared to 92.01% mean recognition performance of traditional approaches like the hidden Markov model and deep belief network. An AI-enabled fog/edge computing approach for fall detection by Vimal et al. (2021) to process binary images of elderly patients in a remote health monitoring setup. The proposed approach has five layers a sensor elder patient body, edge gateway, fog node layer with LoRa connectivity, cloud layer, and application layer for user accessibility. A deep learning convolutional neural networks (CNN) model was used for image processing and compared its performance with support vector machines (SVM) and artificial neural networks (ANN) models. The proposed deep learning model achieved an accuracy of 98% with a minimal processing time of <200 s but with a higher power consumption of >65 decibels.

All the research works discussed in this section have been summarized with their application, algorithms, and technologies used in the RPM system, as shown in Table 4.

## 3.5 | Blockchain monitoring

Virtual technologies like fog computing and edge computing are prone to security and privacy challenges (Aliyu et al., 2021). Hathaliya et al. (2019) proposed a Permissioned blockchain-based healthcare architecture to overcome these challenges. The study focused on integrating decentralized AI with blockchain networks and discussed blockchain applications in healthcare. Blockchain is a shared, decentralized, immutable ledger that connects multiple parties and records transactions. In RPM applications, blockchain technology can secure data transactions between patients and monitoring technologies like cloud, fog, and edge computing. Faruk et al. (2021) proposed an Ethereum-based data repository for RPM electronic health records data management. The data repository enabled secure upload, storage, analysis, retrieval, and transmit patient data according to the patient's instructions. The proposed decentralized blockchain system supports hospitalized patients and outpatients. The cloud computing challenge of interoperability can be addressed using MedHypChain proposed by Kumar and Chand (2021). MedHypChain is a privacy-preserving medical data-sharing system based on Hyperledger Fabric, in which each data transaction is secured via an Identity-based broadcast group encryption scheme. Another interesting patient-centric secured data recording and remote patient monitoring application SynCare was proposed by Pighini et al. (2022). The study focused on interconnecting patients, healthcare professionals, and caregivers, building secure data-sharing channels, and allowing patients to manage their health data. Blockchain architectures are known for their robust security features that record each transaction throughout the system and cannot be altered. The architecture has the disadvantage of high implementations with complex integration and high energy dependence. The research works adapted to blockchain technology in RPM have been outlined in Table 5.

The technology-enabled RPMs are more concentrated on data acquisition and securing the data transmission to different parties involved in RPM. Adopting AI to the RPM architectures empowers the monitoring process with

**TABLE 5** Blockchain architectures

| References | Algorithm | Technology | Advantages | Disadvantages |
| --- | --- | --- | --- | --- |
| Hathaliya et al. (2019)<br>Faruk et al., 2021<br>Kumar and Chand (2021) | AI methods | Blockchain<br>Blockchain, Ethereum<br>Blockchain, Hyperledger Fabric | • Network security at all levels of data collection<br>• Verification and identification of patients<br>• Authorize patients' EHR data | • High energy dependence<br>• Integration complexity<br>• High implementation costs |

Abbreviation: AI, artificial intelligence.

capabilities of prediction and classification of the patient data acquired. Each RPM architecture can be enhanced by adding AI modeling to the data analytics step.

## 4 | AI IN RPM APPLICATIONS

In RPM applications, traditional machine learning and deep learning are common AI methods adopted to detect and predict vital signs and classify patients' physical activities. Malasinghe et al. (2017) present contact and noncontact-based methodologies in RPM. Irrespective of contact or noncontact monitoring systems, all methodologies focus on human vital signs extraction, such as heart rate, pulse rate, respiration rate, blood pressure, and oxygen volume in blood, as the deterioration of these vital signs affects the human health system. Along with vital human signs, the authors reviewed studies on the activity detection of patients like fall detection and mobility-related diseases. The challenges involved mainly include discerning the difference between deliberate quick movements and accidental drops. Apart from wearable devices, the authors reviewed ambiance device-based and vision-based fall detection systems but identified significant problems that remain for contactless monitoring. This section discusses applications of machine learning and deep learning methodologies in RPM. The year-wise distribution of the AI-enabled RPM works discussed in this section are presented in Figure 4.

### 4.1 | Vital signs monitoring

Wearable devices like smartwatches are new technological innovations that continuously track people's vital signs. A system was developed by Bekiri et al. (2020) to monitor the health status of individuals at all times using connected smartwatches. The smartwatches collect patient vital signs and send them to the administrator to analyze for decision-making. The administrator used the SVM model to build a decision model. The results of the patient's status will be informed to the doctors. The machine learning model achieved an accuracy of 90% and a recall is 99%. The proposed system can identify 99% of patients affected by cardiovascular diseases. Shao et al. (2020) also designed an RPM system to detect ECG signals. In that study, a decision tree ensemble classifier was trained using the CatBoost learning kit. The classifier was trained with 20-fold cross-validation and 31 features. Feature importance was extracted from the trained CatBoost model. The top-importance features were used to evaluate the performance based on the feature importance ranking. The CatBoost model processed the 30 s ECG data in 0.5 s and achieved a sensitivity of 99.61%, a specificity of 99.64%, and an accuracy of 99.62% in detecting AF. A novel IoT-based wearable 12-lead ECG SmartVest system based on the SVM model to assess signal quality has achieved an average accuracy of 97.9% and 96.4% for acceptable and unacceptable ECG segments, respectively. Verified the model efficiency to choose good or exclude poor quality ECG segments in the wearable.

ECG monitoring. An SVM model-based ECG telemetry system to monitor cardiac arrhythmia, which processes the ECG signal, was designed by Devi and Kalaivani (2019) to send alerts to a physician in an emergency. Statistical features of ECG signals were combined with dynamic features like heart rate variability (HRV) features from RR intervals to classify cardiac arrhythmia. The SVM classifier model was trained and validated using 10-fold cross-validation. The proposed classification model achieved the effectiveness of 88.9%, 90.8%, and 92.2% for statistical features, HRV features, and statistical and HRV features, respectively.

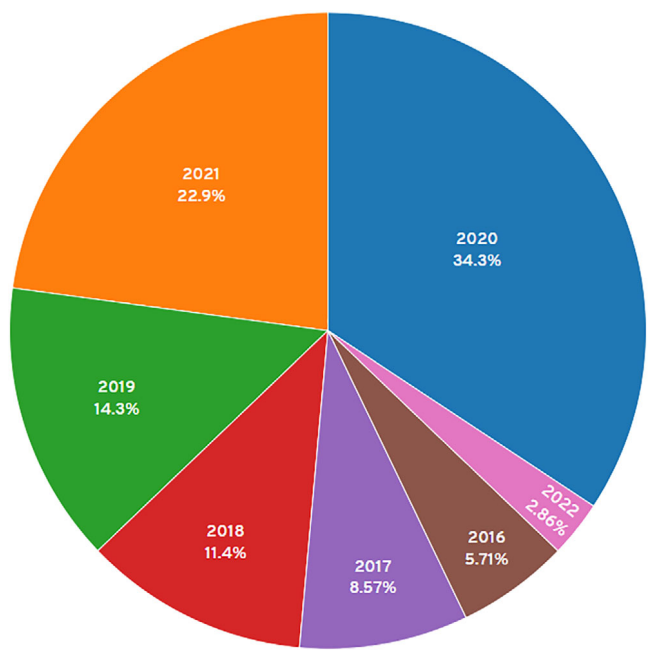

**FIGURE 4** Distribution of artificial intelligence-enabled remote patient monitoring applications in this section

Neto et al. (2017) designed an RPM system with a portable ECG device to assist remote electrocardiographic diagnosis and send the data to cloud service, where an intelligent arrhythmia detector (IDAH-ECG) detected abnormal heartbeats and informed physicians. Here, discrete wavelet transforms feature extraction, and principal component analysis (PCA) dimensionality reduction was performed as part of data preprocessing. Multilayer perceptron (MLP) neural network (MLP) classifier was trained using backpropagation and Gradient Descent techniques. This study implemented 10-fold cross-validation with the Monte Carlo testing scheme. The classifier achieved an average accuracy of 96.48%, a sensitivity of 98.70%, and a specificity of 94.45%. Elola et al. (2019) used deep learning methods for sensor-based pulse recognition from short electrocardiogram (ECG) segments that included a deep convolutional neural network (DCNN), auto-encoder, restricted Boltzmann machine (RBM), and recurrent neural network (RNN). The system was designed to detect a pulse in someone who had a heart attack during resuscitation efforts. Although the deep neural network (DNN) architectures outperformed current methods, pulse detection during this scenario remains an unsolved problem. J. Yang et al. (2020) developed a wireless nonline-of-sight (NLOS) bio-radar device that was used to collect physiological parameters such as heart rate and respiratory rate. The device is portable, contactless, and interference-free. A deep learning LSTM was employed at edge nodes to predict the physiological parameters. The authors used LSTM to predict future short-term respiratory rates and patients' heartbeats based on current data within just a few minutes. W. Qi and Aliverti (2020) proposed a wearable respiratory and activity monitoring system to predict breathing patterns during daily activities based on a novel multimodal fusion architecture, respiratory and exercise parameters and human activity. A hybrid hierarchical classification algorithm combining an LSTM model with a threshold-based approach to classify nine breathing patterns while performing 15 physical activities. The hybrid model achieved an accuracy of 97.22% and outperformed the other models' K-nearest neighbor (KNN), multiclass SVM, and artificial neural network (ANN) in terms of classification. The proposed model outperformed LSTM, bidirectional LSTM (Bi-LSTM), and DCNN with a minimal computational time of 0.0094 s. The research works related to vital signs monitoring with AI discussed are consolidated in Table 6.

## 4.2 | Physical activities monitoring

Pan et al. (2020) designed a fall detection system for older people based on multisensor fusion with multiple three-axis acceleration sensors placed on the waist. In this study, SVM and random forest (RF) algorithms were implemented on the dataset with 100 healthy young volunteers simulating falls and daily activities to compare their recognition time and recognition rate. The authors state that the model's accuracy is based on a large amount of valid data, but because SVM has fewer training and recognition times, it may be better suited to this task. Hsieh et al. (2021) proposed a novel multiphase falls identification algorithm combining fragment modification algorithm and machine learning techniques to identify prefall, free-fall, impact, resting, and recovery phases. The fragment modification algorithm adopts rule-based fall identification and five machine learning techniques, SVM, KNN, naive Bayes, decision tree, and adaptive boosting to identify the five phases. Out of the five models, the KNN algorithm achieved the best performance with an accuracy of 90.28%, sensitivity of 82.17%, precision of 85.74%, and Jaccard coefficient of 73.51%. The authors intend to further develop their model with real-world data and a greater range and type of falls. Y. Wang and Zheng (2018) designed a framework for an RPM system to monitor human activities and movement based on a signal reflection model. This framework detected the presence of human activities by analyzing the RSSI patterns from an RFID tag array and segmented phase values using the variance of phase readings, which were used as an indicator for activity segmentation. Six machine learning classifiers RF, multilayer Perceptron-based Neural Network, Decision Tree, SVM, Naive Bayes, and Quadratic Discriminant Analysis, were trained to classify activities raise a hand, drop hand, walk, sit, stand, fall, rotation, get-up, and non-activity. The experiment results show that TACT is robust under different experimental settings and can achieve an average recognition precision of up to 93.5%.

**TABLE 6** Vital signs monitoring with artificial intelligence

| Applications | Algorithm | Technology | References |
| --- | --- | --- | --- |
| Vital signs monitoring | SVM Model | Smartwatches, smart vest, ECG telemetry | Bekiri et al. (2020), Shao et al. (2020), Devi and Kalaivani (2019) |
| | CNN, LSTM, DCNN, DNN, RNN, ANN auto-encoder | Potable ECG device, sensors, radar device | Neto et al. (2017), Elola et al. (2019), J. Yang et al. (2020), W. Qi and Aliverti (2020) |

Salah et al. (2022) designed a resource-constrained microcontroller at the edge of the network using a wearable accelerometer to overcome issues such as latency, high power consumption and poor performance in areas with unstable internet. The authors designed three layers edge layer, fog layer, and cloud layer to collect, analyze, and transmit to an IoT gateway via long-range communication technology. Five AI models, KNN, SVM, LSTM, and CNN, were trained to detect falls. The LSTM identified falls from daily activities with high accuracy of 96.78%, while sensitivity and specificity were 97.87%, and 95.21%, respectively. S. Yu et al. (2021) proposed a computational method with a Hierarchical Attention-based Convolutional Neural Network (HACNN) model to detect falls based on wearable sensor data. The novel deep learning model integrated a hierarchical attention mechanism into a CNN model and added two attention layers beyond CNN to interpret which part of the sensor data contributed to the decision of fall or nonfall made by the system. The CNN model outperformed deep learning models like CNN, LSTM, CNN-LSTM, MLP, and HALSTM. Accuracy depended on the two data sets used and their static nature.

To overcome the limitations of existing elderly fall detection methods requiring specialized hardware or invading people's daily lives, Zhu et al. (2017) presented the design and implementation of a motion detection system based on passive radio frequency identification tags. The received signal strength indicator (RSSI) value and Doppler frequency value impacted by static, regular action, sudden falls, elderly movements, and fall actions were estimated. Wavelet transform was implemented for the signal preprocessing, and the machine learning algorithm SVM was adopted to classify the actions into fall detection or other actions. RFID technology could track their motion and fall detection without any hindrance to the daily activities of elderly people. Gesture recognition or motion detection has gained attention to enhance the user experience for human-computer interaction. An application can be used in healthcare to recognize patient gestures or motions in-home or in the hospital using a device-free system. Z. Wang et al. (2019) proposed RF-finger, a device-free system based on Commercial-Off-The-Shelf (COTS) RFID, which leverages a tag array on a letter-size paper to sense the fine-grained finger movements performed in front of the paper presented. Machine learning algorithms were implemented, such as the KNN model to pinpoint the finger position and the CNN model to identify the multitouch gestures based on reflective images. Both the machine learning algorithms yielded 88% and 92% accuracy for finger tracking and multitouch gesture recognition, respectively. Estimate the correlation between RF phase values and human activities by modeling intrinsic characteristics of signal reflection in contact-free scenarios. The research works related to human activity recognition and fall detection are presented in Table 7.

TABLE 7  Human activity recognition with artificial intelligence

| Applications | Algorithm | Technology | References |
| --- | --- | --- | --- |
| Physical Activities Monitoring | SVM, RF, KNN, naive Bayes, decision tree, adaptive boosting | Sensors, RFID Tags | Pan et al. (2020), Hsieh et al. (2021), Y. Wang and Zheng (2018), Zhu et al. (2017); Z. Wang et al. (2019) |
|  | KNN, SVM, LSTM, CNN, HACNN | Wearable accelerometer, cloud, fog, edge | Salah et al. (2022), S. Yu et al. (2021) |

## 4.3 | Chronic disease monitoring

### 4.3.1 | Diabetes monitoring

Mujumdar and Vaidehi (2019) proposed a diabetes prediction model to classify diabetes which includes external factors responsible other than regular factors like glucose, body mass index (BMI), age, insulin, and so on. In this study, machine learning algorithms, including Support Vector Classifier, Decision Tree classifier, Extra Tree Classifier, Ada Boost algorithm, Neural Networks, RF Classifier, Linear Discriminant Analysis algorithm, Logistic Regression, K-Nearest Neighbor, Gaussian Naïve Bayes, Bagging algorithm, and Gradient Boost Classifier was implemented for diabetes prediction. All the models were trained and evaluated using a confusion matrix and classification report. Out of all, the logistic regression model was able to classify diabetic and nondiabetic with an accuracy of 96%. To continuously evaluate diabetic patients' data such as sugar level, sleep time, heart pulse, food intake, and exercise collected through sensors. Analyze the data using neural networks and classify the health risk status into four modes: low, medium, high, and extreme (Efat et al., 2020). Based on national physical examination, the risk factors for type II diabetes mellitus (T2DM) were computed using machine learning algorithms. A logistic regression model was implemented on physical measurement and a questionnaire. The 14 risk factors selected in logistics regression were combined and implemented with tree-based machine learning algorithms like decision tree, RF, AdaBoost, and XGBoost. Out of 4 algorithms, XGBoost had achieved an accuracy of 90.6%, precision 91.0%, recall 90.2%, F1 score 90.6% and AUC 96.8%. XGBoost model was used to output feature importance scores. BMI was the most important feature, followed by age, waist circumference, systolic pressure, ethnicity, smoking amount, fatty liver, hypertension, physical activity, drinking status, dietary ratio (meat to vegetables), drink amount, smoking status, and diet habit (oil-loving) (Xue et al., 2020). Several ML techniques could classify different states of diabetic patients with high accuracy.

### 4.3.2 | Mental health monitoring

Mental health illness is one of the most underestimated human states, which shortens the life span by 10–20 years (McGinty et al., 2021). It would be difficult to manually monitor people with mental health illnesses such as schizophrenia, bipolar disorder, major depressive disorder, and suicidal tendency. Thieme et al. (2020) conducted a systematic review on how implementing machine learning can assist in detecting, diagnosing, and treating mental health problems. Machine learning techniques can offer new routes for learning patterns of human behavior. It helps in identifying mental health symptoms and risk factors. Also, it assists in predicting disease progression and personalizing and optimizing therapies. Advances in machine learning will attempt to predict suicide based on the analysis of relevant data and inform clinical practice. Adamou et al. (2019) proposed a text-mining approach to support risk assessment. Latent Dirichlet allocation (LDA) is a special case of topic modeling for natural language processing. The technology was used to process different types of information like demographics, appointments, progress notes, comprehensive assessments, referrals, and Inpatient stays. Statistically equivalent signatures (SES) mechanism for feature selection. In this study, support vector regression (SVR), RF, and linear ridge regression (RLR) models were implemented along with K-fold cross-validation. The approach achieved the highest area under curve (AUC) value of 0.705. Continuous monitoring would enable the record of abnormal vital sign measurements, and ML techniques have the potential to analyze the data to detect underlying data patterns to take appropriate treatment steps. Diabetes and Mental Health monitoring discussed in this section are outlined in Table 8.

**TABLE 8** Chronic disease monitoring with artificial intelligence

| Applications | Algorithm | Technology | References |
| --- | --- | --- | --- |
| Diabetes monitoring | Decision Tree, RF, AdaBoost, XGBoost, RF, KNN, SVM | EHRs, Sensors | Mujumdar and Vaidehi (2019), Efat et al. (2020), Xue et al. (2020) |
| Mental health monitoring | LDA, SVM, RF, RLR, SES | Social media text, demographics | Thieme et al. (2020), Adamou et al. (2019), McGinty et al. (2021) |

## 4.4 | Emergency monitoring

### 4.4.1 | Emergency department

RPMs Decision-making for emergency department patients using machine learning techniques would have helped to improve existing methods. Taylor et al. (2016) compared clinical decision rule to a machine learning approach for predicting in-hospital mortality of patients with sepsis. In this study, machine learning techniques were used to extract a large number of variables through existing emergency department clinical records to predict patient outcomes and facilitate automation and deployment within clinical decision support systems. Patients visiting the emergency department visits were randomly partitioned into an 80%/20% split for training and validation. 500 clinical variables were extracted from the real-time clinical records of four hospitals using an RF model to predict in-hospital mortality. Later, the RF model was compared to the classification and regression tree (CART) and logistic regression models. The random forest model AUC was statistically different from all other models ($p \leq 0.003$ for all comparisons). Kong et al. (2016) designed a decision tool based on rule-based inference methodology using the evidential reasoning approach (RIMER) that was developed and validated to predict trauma outcomes. It helps physicians to predict in-hospital death and intensive care unit (ICU) admission among trauma patients in emergency departments. The prediction performance of the RIMER was compared to logistic regression analysis, support vector machine, artificial neural network, SVM models, and ANN models. Five-fold cross-validation was implemented, and the AUCs of RIMER, logistic regression, SVM, and ANN are 0.952, 0.885, 0.821, and 0.790, respectively. The results show that the RIMER model performs the best. The machine learning techniques could classify near-term mortality based on vital signs analysis in emergency department patients.

The machine learning approach is able to incorporate heart rate variability (HRV) for intensive monitoring, resuscitation facilities, and early intervention for critically ill patients in the emergency department by comparing the area under the curve, sensitivity, and specificity with the modified early warning score (MEWS). In a study (Oh et al., 2018), HRV parameters were generated from a 5-min electrocardiogram (ECG) recording incorporated with age and vital signs to generate the ML score for each patient. The area under the receiver operating characteristic curve (AUROC) for ML scores in predicting cardiac arrest within 72 h is 0.781, compared with 0.680 for MEWS. For in-hospital deaths, the area under the curve for ML score is 0.741, compared with 0.693 for MEWS. A cut-off machine learning score $\geq 60$ predicted cardiac arrests with a sensitivity of 84.1%, specificity of 72.3%, and negative predictive value of 98.8%. A cut-off MEWS $\geq 3$ predicted cardiac arrest with a sensitivity of 74.4%, a specificity of 54.2%, and a negative predictive value of 97.8% (Blasiak et al., 2020; Ong et al., 2012). Based on the results, machine learning scores were more accurate than the traditional MEWS in predicting cardiac arrest within 72 h.

### 4.4.2 | RPMs in the ICU

To predict near-term mortality in patients hospitalized with cirrhosis (Antunes et al., 2017), two machine learning approaches (i) logistic regression and (ii) LSTM neural network on medical record entries of 500 patients staying ICU and compared them (Xia et al., 2019). In total, 20 features, such as pulse, respiratory rate, systolic, and diastolic blood pressure, were used for training the algorithm. The machine learning models outperformed the clinical decision tool, a mathematical Chronic Liver Failure (CLIF) Score. A logistic regression model achieved an AUC of 0.80, the RNN-LSTM model achieved an AUC of 0.77, and CLIF achieved an AUC of 0.72 (Harrison et al., 2018). A patient-specific model could analyze vital signs based on historical data. Colopy et al. (2018) proposed Gaussian process regression (GPR) to provide flexible, personalized models of time series of patient vital signs. This study uses a method to build GP models with varying complexity and regularization using different hyperparameters on a patient-specific level to forecast robust, vital signs. The authors used a random search algorithm to search for patient-specific parameters. Bayesian optimization methods were implemented to accommodate any plausible parameterization in the patient population. Patient-specific parameter optimization using machine learning techniques is the most advanced level of RPM. This helps to build patient-specific models and break down a patient's health status to the lowest level. The Table 9 consolidates the research works related to emergency and ICU patients monitoring.

**TABLE 9**  Emergency and intensive care unit patients monitoring with artificial intelligence

| Applications | Algorithm | Technology | References |
| --- | --- | --- | --- |
| Emergency monitoring | RF, RIMER, LSTM, GPR, Logistics Regression, CLIF | EHRs, ECG, RIMER | Taylor et al. (2016), Kong et al. (2016), Oh et al. (2018), Ong et al. (2012), Blasiak et al. (2020), Antunes et al. (2017), (Xia et al. (2019), (Harrison et al. (2018), Colopy et al. (2018) |

**TABLE 10**  Facial and emotions recognition with artificial intelligence

| Applications | Algorithm | Technology | References |
| --- | --- | --- | --- |
| Facial and emotions monitoring | LSTM, SVM | Image Processing Thermal sensors IoT devices | Mahesh et al. (2021), Chowdary et al. (2021), Zainuddin et al. (2020) |
|  |  | RFID signals | Q. Xu et al. (2020) |

## 4.5 | Facial and emotions recognition

AI can classify the patient's emotions based on patient face recognition. A smart integrated patient monitoring system was proposed by Mahesh et al. (2021) to detect patients' emotional states and heartbeat levels through face recognition algorithms, heartbeat, and temperature sensors. Their RPM system presented the emotional data of the patients using face recognition algorithms such as image preprocessing, feature extraction, and classification. The facial emotional recognition model is able to identify seven emotions: Anger, Happy, Sad, Neutral, Surprise, Disgust, and Fear. Based on the heart rate sensor and thermal sensor, patients' vital signs were measured with an interval of 5 s. Zainuddin et al. (2020) used IoT technology to communicate facial emotions and vital signs to hospitals. Similarly, Chowdary et al. (2021) designed an RPM system based on deep learning-based facial emotion recognition to overcome problems associated with mutual optimization of feature extraction and classification.

An experimental study to recognize user emotions of users with commercial RFID devices. Q. Xu et al. (2020) designed an RPM system using an emotion recognition framework that first extracts respiration-based features and heartbeat-based features from RFID signals. The extracted features were used in training a classifier that a user's different emotions. In this study, the respiration rate was separated by using filters, whereas the heartbeat signal was retrieved by suppressing the respiratory signal and improving the signal-to-noise ratio. Using their framework, a 2D emotional model divided emotions into four states: joy, pleasure, anger, and sadness. An SVM model was able to classify the four emotion states with an accuracy of 80.65%, 61.29%, 83.87%, and 74.19%, respectively (Q. Xu et al., 2020). The research community developed RPM systems using AI for facial and emotions recognition with technologies, as shown in Table 10.

## 5 | AI IMPACT ON RPM

### 5.1 | Early detection of patient deterioration

Early detection of vital signs deterioration is key to the timely invention and avoiding clinical deterioration in acutely ill patients in hospitals. Traditional patient monitoring is to report individual vital signs of patients, which state their current clinical status. For example, vital signs such as temperature, pulse, respiratory rate, and mean arterial pressure (MAP) are considered continuous predictors for emergency department patients (Asiimwe et al., 2020). New patient monitoring algorithms analyze multiple features from physiological signals. This produces a predictive or prognostic index that measures a specific critical health event or physiological instability (Helman et al., 2022). Posthuma et al. (2020) presented a case series where wireless remote vital signs monitoring systems on surgical wards could reduce the time to detect deteriorating patients. As part of this study, nursing staff found the systems somewhat useful, but still required clinical judgment to assess the patient. They noted that there are still no set standards or guidelines for implementing these types of systems, and the task remains for clinicians to judge which system best meets their needs.

Kellett and Sebat (2017) further elaborated on the need for clinicians to place more importance on regular and accurate recording of vital signs. The authors noted that there is currently no agreement on how often vital signs need to be recorded and that most hospital wards use periodic, manual observation of vital signs. Kellett also highlighted the need for continuous patient monitoring and emphasized how vital this is to predict the onset of abnormal events.

Current approaches by clinicians for early prediction of patient deterioration can be estimated using manually calculated screening metrics called early warning scores (EWS) (Garca-del Valle et al., 2021; Vinegar & Kwong, 2021). Downey et al. (2018) demonstrated that although EWS systems have excellent predictive values, they are limited by their intermittent nature. Until recently, continuous vital signs monitoring was limited to intensive care units that require high staff-to-patient ratios. For example, Alshwaheen et al. (2021) proposed a novel framework of patient deterioration prediction in ICUs based on LSTM-RNNs. The model acquired a significantly better classification performance than the traditional method and could predict deterioration 1 h before onset. Muralitharan et al. (2020) further showed that machine learning based EWS could be applied to a range of acute general medical and surgical wards, including ambulatory and home care settings, and still perform better and with greater accuracy than the traditional manual methods.

Although many studies have focused on the prediction of health outcomes, da Silva et al. (2021) predicted future deteriorating vital signs based on applying RNNs and LSTM to historical data from electronic medical records (EMR). These predicted vital signs were then applied to a clinical prognostic tool that used a combination of laboratory results with vital signs for early diagnosis of worsening health status, with an accuracy of 80%.

Transparency and explainability are essential elements for AI models if they are going to be acceptable to clinicians. Lauritsen et al. (2020) proposed an explainable AI EWS (xAI-EWS) system for the early detection of acute critical illness. The xAI-EWS was composed of a temporal convolutional network (TCN) prediction module and a deep Taylor decomposition (DTD) explanation module tailored to temporal explanations. Clinical experts evaluated the system based on three emergency medicine cases: sepsis, acute kidney injury (AKI), and acute lung injury (ALI). The system facilitated trust in the predictive capability by giving clinicians insights into the internal mechanics of the model without any deep technical knowledge of the mechanisms behind it.

## 5.2 | Personalized monitoring

Conventional diagnoses of diseases and treatments from doctors are based on population averages and do not consider the individual variability of patients to treatments (G. Chen, Xiao, et al., 2021). The IoT-enabled RPM architecture with cloud computing discussed in previous sections combines patients' data for AI modeling. In contemporary settings, patient-centric or personalized monitoring is critical, particularly for chronic diseases like mental health disorders, diabetes, and so on. Personalized monitoring can be carried out with distributed networks like fog and edge computing, where an edge network is set up for a set of IoT devices on a patient. Mukherjee et al. (2020) proposed an edge-fog-cloud framework for personalized health monitoring to predict patient mobility and advice nearby healthcare centers in case of emergency. However, the data acquired from an IoT platform has to leave the devices and be merged into a centralized cloud server for data analytics. This raises privacy and security concerns about patients' health data. Moreover, it demands huge technological resources and power consumption.

A Federated learning framework in AI methods developed by Google could overcome the challenges by training an AI model across multiple decentralized edge devices with local data available at each patient without exchanging or merging them. The local model weights are aggregated and passed to a cloud server. The aggregated model weights are used to train a robust global AI model. In this decentralized framework, the patient data will not leave their device and ensure data privacy. The robust global model can be passed to local models for better prediction or classification results on local data. The research community has widely adopted the approach for IoT applications. Zheng et al. (2021)

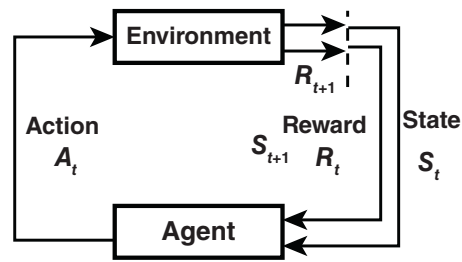

**FIGURE 5** Generic reinforcement learning mechanism

proposed a federated transfer learning mechanism for the internet of medical things (IoMT) healthcare. Nguyen et al. (2023) discussed types of federated learning frameworks for smart healthcare, benefits, requirements, federated learning applications in applications, trends, and challenges. Wu et al. (2022) proposed the FedHome framework, a novel cloud-edge-based federated learning framework for in-home health monitoring to training local models. In that study, a generative convolutional autoencoder (GCAE) was designed to process imbalanced and nonidentical distribution data and to achieve accurate results in personalized health monitoring. The proposed approach outperforms baseline models with an accuracy of 95.87% and 95.41% for balanced data and imbalanced data, respectively. In personalized monitoring, physical activity classification task was performed with FedHealth by Y. Chen et al. (2020). The federated learning framework proposed in the study is based on data aggregation and builds personalized models with transfer learning. FedHealth framework was evaluated with two classification problems. One is to classify physical activities and outperform the baseline models with an accuracy of 99.4%. The second one is to classify Parkinson's disease patient's arms droop and postural tremor test and achieved an average accuracy of 84.3% and 74.9%, respectively. Similarly, Shaik, Tao, Higgins, Gururajan, et al. (2022); Shaik, Tao, Higgins, Xie, et al. (2022) proposed a heterogeneous FedStack framework to support diverse architectural local models at patients-end to classify their physical activities and build a robust global model based on predictions of the local models.

## 5.3 | Adaptive learning

Reinforcement learning, a subset of AI, possess the ability to make a sequence of decision with its reward-driven behavior. The machine learning approach learns to achieve a goal in a potentially uncertain, complex environment. It can employ trial and error to solve a problem and get either rewards or penalties for the steps it executes (C. Yu et al., 2023). In the reinforcement learning approach, a learning agent is deployed in an environment without any prior information or knowledge. The agent has to learn the patterns based on their experience. To transit from the current state ($S_t$) at time $t$ to the next state ($S_{t+1}$) at time step $t + 1$, an action ($A_t$) is taken as shown in Figure 5. For these actions, a predefined reward policy is designed. If the actions chosen is following the policy, the agent gets rewarded ($R_t$), otherwise penalized. With the sequential decision-making capability, different reinforcement learning schemes are applied to diverse dynamic treatment regimes (Laber et al., 2014) like chronic diseases, mental health diseases, and infectious diseases which need a sequence of decision rules to determine a course of action to suggest treatment type, drug dosage, or re-examination timing. C. Yu et al. (2023) surveyed the applications of reinforcement learning in healthcare. The study has covered treatment strategies built on reinforcement to treat chronic diseases, cancer, diabetes, anemia, HIV, and several common mental illnesses. I. Y. Chen, Joshi, et al. (2021) considered a clinician (learning agent) who monitors the patient (environment) via actions like ventilation and observing the changes in the patient's state (environment) to achieve a goal to discharge the patient successfully. This study provides a practical understanding of the reinforcement learning approach in healthcare. Watts et al. (2020) developed a model to prescribe the timing and dosage of medications using wearable sensors in real-time and deep reinforcement learning. Similarly, Naeem et al. (2021) proposed an intelligent system that relies on algorithms of both Reinforcement Learning and Deep Learning to maximize the successful completion of the patient taking the right pill. Just-in-Time Adaptive Interventions (JITAIs) are another healthcare applications which needs timely intervention to provide the right amount of support to patients at right time. This can be achieved by adaptive learning of dynamic health changes in a patient (Nahum-Shani et al., 2017). Wang, Zhang, et al. (2021) adopted reinforcement learning in a data-driven approach for mobile healthcare user and optimize intervention strategies in their context. Similarly, Gönül et al. (2021) proposed a reinforcement learning mechanism to personalize digital adaptive interventions as mobile notifications to the user in coping their health problems. The authors deployed two models, intervention selection and opportune moment identification. With respect to type and frequency, the intervention selection model adopts the intervention delivery. The opportune moment identification is to detect the most opportune moments to intervene. Table 11 presents research works of AI which can transform healthcare applications with advanced mechanism such as reinforcement learning and federated learning.

**TABLE 11** Artificial intelligence (AI) impact on remote patient monitoring systems

| AI impact | Algorithms/technology | Applications | References |
| --- | --- | --- | --- |
| Early detection of patient deterioration | explainable AI EWS, LSTM, TCN, DTD | • Continuous monitoring of emergency patients, sepsis, acute kidney injury, acute lung injury<br>• Early diagnosis with Early Warning Scores and Predictive Prognostic Index | Asiimwe et al. (2020), Helman et al. (2022), Posthuma et al. (2020), Kellett and Sebat (2017), Garca-del Valle et al. (2021), Vinegar and Kwong (2021), Downey et al. (2018), Alshwaheen et al. (2021), da Silva et al. (2021), Lauritsen et al. (2020) |
| Personalized monitoring | Fog, Edge, IoMT, Cloud GCAE, FedHealth, FedStack | • Enable personalized monitoring with decentralized learning<br>• Overcome data privacy issues | G. Chen, Xiao, et al. (2021), Mukherjee et al. (2020), Zheng et al. (2021), Nguyen et al. (2023), Wu et al. (2022), Y. Chen et al. (2020), Shaik, Tao, Higgins, Gururajan, et al. (2022); Shaik, Tao, Higgins, Xie, et al. (2022) |
| Adaptive learning | DRL, A2C, DQN | • Learn patient behavior patterns<br>• Dynamic treatment regimes<br>• Just-in-time-adaptive-interventions<br>• Sequential decision making tasks | C. Yu et al. (2023), Laber et al. (2014), I. Y. Chen, Joshi, et al. (2021), Watts et al. (2020), Naeem et al. (2021), Nahum-Shani et al. (2017), Wang, Zhang, et al. (2021), Gönül et al. (2021) |

# 6 | CHALLENGES AND TRENDS OF AI IN RPM

Implementing a technology-enabled patient monitoring system would require hospital staff support and their views. Ede et al. (2021) did a qualitative study to explore staff expectations of wireless noncontact patient vital signs monitoring, their perception of the utilization of the technology in the ICU, patients, and relative response to introducing the technology. Nine nurses with a median duration experience of 2 years in ICU were interviewed on five different themes such as ICU staff perceptions of the patient and relative monitoring experiences relating to current wired monitoring and expectations of noncontact monitoring, staff expectation of continuous monitoring in ICU, troubleshooting, the hierarchy of monitoring and consensus of trust. Although AI can transform healthcare with its potential to analyze, predict and classify data efficiently, there remains a hesitancy to adopt the technology (Meskó et al., 2017). This section discusses challenges in adopting AI to remote monitoring systems for vital signs precision and activity recognition. Initiatives to overcome the challenges are also presented.

## 6.1 | AI or ML explainability

The first and foremost challenge is the difficulty associated with interpreting the results generated by an AI or ML model. Current models are better than humans at interpreting complex data and predicting outcomes but lack the capacity to demonstrate how these conclusions were reached or if there were any weaknesses in the algorithm applied by the ML model. This is one of the most challenging barriers for healthcare professionals to adopt AI or machine learning methodologies (Mohanty & Mishra, 2022). Most of the machine learning models, such as neural networks, SVM, and so on are black-box models. These models cannot elaborate their results and provide cause-and-effect relationships between predictor variables and target variables (G. Yang et al., 2022). AI or ML can be adaptable only when interpretable structures and results for healthcare professionals (Sagi & Rokach, 2020).

Sensitivity is one of the promising methods which can explain the cause-and-effect relationship between the input and output variables of a trained neural network. Tree-based methods will not expect a parametric relationship between input and output variables. Both classification and regression trees have been shown to provide interpretable structures that would support clinical practice in decision-making (Jovanovic et al., 2016; Sagi & Rokach, 2020). Other than tree-based methods, pattern-based classification, Naive Bayes, knowledge-based algorithms, logistic regression, and fuzzy models can produce interpretable structures (Shouval et al., 2017).

SHapley Additive exPlanations (SHAP) (Lundberg & Lee, 2017), a method based on cooperative game theory (Shapley, 1953), can increase the transparency and explainability of AI Models. In this method, the impact or contribution of input to the prediction output is represented with Shapley values and the values are calculated for each input feature. The Shapley values of the prediction model were extracted in two forms: a global perspective of factors that required special attention in overall prediction, and a local perspective of each feature in a single prediction. Linardatos et al. (2020) reviewed machine learning interpretability methods. The study reported different scopes of interpretability for deep learning models. It includes gradients explanation technique, integrated gradients, gradient-weighted class activation mapping, DeepLIFT algorithm, deconvolution, and guided back-propagation. Raza et al. (2022) proposed a framework for accurate and efficient personal healthcare using federated transfer learning and an explainable AI (ExAI) model in EEG signal classification. Khodabandehloo et al. (2021) proposed a flexible AI system HealthXAI to predict early symptoms of early decline in smart homes. The anomaly level of behavior was computed based on the anomaly feature vector. The authors built a dashboard to allow clinicians to inspect anomalies, scores, and their automatically generated natural language explanations. Trends in Explainable AI are sensitivity and Shapley values and the research works exploring these trends in Table 12.

## 6.2 | Privacy

Considering the black-box nature of deep neural networks, it is impossible to predict what neural networks learn from data. The problem with this is that they might unintentionally learn features that discriminate against user information. This increases the risk of information disclosure. Iwasawa et al. (2017) analyzed the features learned by conventional deep neural networks when applied to data of wearable to confirm this phenomenon. A simple logistic regressor could achieve a high user classification accuracy of 84.7% when using the CNN features extracted from basic activity signals. The same classifier could only obtain 35.2% user classification accuracy on raw sensor data. This reveals the privacy leakage potentials of a deep learning model originally used for human activity recognition (K. Chen, Zhang, et al., 2021).

In this study (K. Chen, Zhang, et al., 2021), data transformation and data perturbation techniques, were suggested that could be used to overcome privacy issues with machine learning algorithms. User adversarial neural networks were proposed to integrate an adversarial loss with the standard activity classification loss to minimize the user identification accuracy. However, the adversarial loss technique has a limitation to protecting only private information, such as user identity and gender. To protect all sensitive user identity information, the raw sensor signals were viewed from two perspectives, style and content. D. Zhang et al. (2019) proposed to transform raw sensor data to have the "content" unchanged, but the "style" is similar to random noises. For data perturbation, a deep private auto-encoder (dPA) was proposed by Gati et al. (2021) to perturb the objective functions of the traditional deep auto-encoder to enforce $\epsilon$-differential privacy. In addition to the privacy preservation in feature extraction layers, a $\epsilon$-differential privacy preserving softmax layer was also developed for either classification or prediction. The blockchain technology discussed in Section 3 of this study is one of the trends the research community is adopting to overcome the privacy issue. Hossein et al. (2019) proposed a blockchain-based architecture for e-health applications in which users' data privacy is maintained using features like immutability and anonymity. Ul Hassan et al. (2020) adopted the differential privacy strategy in data perturbation and protect the data in the blockchain. The authors integrated the differential privacy issues in each layer of the blockchain. Another trending AI mechanism, Federated Learning is being adopted for its capacity to collaborate learning and maintain data privacy. The federated learning approach can maintain data privacy by allowing local clients to share only their local AI model parameters, not private data. Singh et al. (2021) combined blockchain and federated learning to propose a secure architecture for privacy-preserving in smart healthcare. The authors take advantage of federated learning features and send only model parameters to the cloud. Data perturbation, blockchain technology and federated learning techniques are being widely adopted to overcome patient privacy and data leakage in healthcare applications as shown in Table 13.

TABLE 12  Trends in explainable artificial intelligence (AI)

| Challenge | Trends | References |
| --- | --- | --- |
| Explainable AI or ML | Sensitivity | Sagi and Rokach (2020), Jovanovic et al. (2016), (Shouval et al. (2017) |
|  | Shapley values | Lundberg and Lee (2017), Shapley (1953), Linardatos et al. (2020), Raza et al. (2022), Khodabandehloo et al. (2021) |

TABLE 13  Trends in protecting privacy

| Challenge | Trends | References |
| --- | --- | --- |
| Privacy | Differential privacy—Data perturbation | K. Chen, Zhang, et al. (2021), Iwasawa et al. (2017), Gati et al. (2021), D. Zhang et al. (2019), |
|  | Blockchain technology | Hossein et al. (2019), Ul Hassan et al. (2020) |
|  | Federated Learning | Singh et al. (2021), Shaik, Tao, Higgins, Gururajan, et al. (2022); Shaik, Tao, Higgins, Xie, et al. (2022), Y. Chen et al. (2020) |

## 6.3 | Uncertainty

There are different uncertainties, such as the data acquisition process, deep neural networks (DNN) building process, and modeling results in adopting AI methodologies to healthcare applications (Gawlikowski et al., 2021). Data acquisition plays a vital role in RPM systems. Still, error, noise in measurement systems, and variability in real-world situations cause uncertainty. While building and training the DNN model with the acquired data would lead to uncertainty in the model structure and training procedure due to a large number of hyperparameters in DNN. The former two uncertainties would lead to uncertainty in the modeling results can be split into data uncertainty (aleatoric uncertainty) and model uncertainty (epistemic uncertainty) (Hüllermeier & Waegeman, 2021). Uncertainty quantification (UQ) can reduce the impact of uncertainties during both the optimization and decision-making process. Abdar et al. (2021) surveyed the research community's work on quantifying uncertainty in machine learning and deep learning models. The review article discussed ensemble techniques and Bayesian techniques like Bayesian deep learning (BDL) (H. Wang & Yeung, 2016) and Bayesian NNs (BNNs) (K. C. Wang et al., 2018) to address the reliability issue of the deep learning models and can interpret their hyperparameters. Begoli et al. (2019) also discussed the need for UQ in machine learning-assisted medical decision-making. The authors discussed four overlapping groups of challenges in UQ especially deep learning models being used in medical applications. The absence of theory in healthcare research is one of the challenges, which means without a fundamental mathematical model, the research is bound to assumptions. The second challenge is the absence of casual models due to limited conclusions from DL models. Sensitivity due to imperfect real-world data while quantifying the uncertainty. The last challenge discussed was computation expense due to deep learning training and re-computation or re-evaluation, causing additional burdens.

## 6.4 | Signal processing

Most signal processing issues remained with noninvasive RPMs that did not touch the patient. Information system infrastructure like RFID reader-antennas was able to retrieve data from RFID tags placed on different areas of the patients. However, transforming the tags' data into vital signs was a challenging task comprising noise (He et al., 2017; Q. Xu et al., 2020). Environmental noise obscured respiration and heartbeat signals in these device-free scenarios. RFID devices utilize a frequency hopping spread spectrum in many countries and regions, causing a discontinuous phase stream. The signal fluctuation caused by intense emotions can overwhelm the respiration and heartbeat signals, resulting in errors in signal extraction (Hou et al., 2017; Zhao et al., 2018). Signal processing challenges could be handled by taking advantage of frequency differences in vital signs and noisy data. It is evident that the double parameter of the least mean square (LMS) (He et al., 2017) can extract a respiration signal with a fundamental frequency (H. Wang et al., 2016; X. Wang et al., 2017). Contact-less respiration and heartbeat monitoring (CRH) systems (Q. Xu et al., 2020; Zhao et al., 2018) that were designed to extract vital signs used smoothing, filtering on raw measurements, and used an intense motion detector system to extract the coarse-grained signal. This was further processed to extract respiratory and heartbeat signals. Noninvasive RPM systems have also used smoothing, unwrapping, interpolation, and Fourier transform techniques to extract breathing and heartbeat signals (Hou et al., 2017). He et al. (2017) applied a frequency-modulated continuous wave (FMCW) radar to monitor vital signs for multihuman targets. The data was collected through the chest wall, with periodic vibration to record respiratory and heart rates. The study proposed a vital signal separation method that could obtain accurate respiration and heartbeat signals using a novel double parameter, the least mean square (LMS) filter. The respiration signal was extracted at the fundamental frequency, and the heartbeat signal from the mixed physiological signal was based on the double-parameter LMS filter. Frequency differences can help in signal processing by adopting techniques such as smoothing, interpolation, Fourier transforms, and frequency filters as shown in Table 14.

## 6.5 | Imbalanced dataset

An imbalanced dataset is a common challenge in AI or ML for data scientists, as it can lead to bias in decision-making. In the supervised machine learning technique, class-imbalanced datasets could affect the predictive ability of the model (Gao et al., 2021). An imbalance in classification categories of a dataset where more samples are from one class is called a majority class, with the other type called a minority class. Conventional machine learning algorithms tend to predict the majority class while ignoring the minority class (Chen, Zhang, et al., 2021; Kaieski et al., 2020). The process could be either using under-sampling techniques like EasyEnsemble and BalanceCascade (Choudhary & Shukla, 2021) to reduce majority class samples or using an over-sampling technique like Synthetic Minority Oversampling Technique (SMOTE) (Hambali & Gbolagade, 2016) to reproduce minority class samples (Alotaibi & Sasi, 2016). Either of these two approaches would adequately deal with class imbalance.

Wang, Yao, and Chen (2021) proposed a long-tail data processing, undersampling-clustering-oversampling algorithm, for heart rate prediction in stroke patients. The authors use a randomly undersampling technique on majority labels and K-Means clustering on minority before applying SMOTE technique on the combined dataset. Kumar et al. (2022) performed a review of class-imbalanced learning situations with six machine-learning classifiers on five imbalanced clinical datasets. The authors explored seven different label balancing techniques such as SMOTE, SVM-SMOTE, ADASYN, Undersampling, Random Oversampling, SMOTETOMEK, and SMOTEEN. Out of all techniques, SMOTEEN with the KNN model achieved the highest accuracy, recall, precision, and F1 score.

Evaluation metrics also play a critical role in addressing bias in class imbalance problems. AI model evaluation results can be misleading. For example, in a multilabel classification problem, considering the overall accuracy of an AI model could show the model's performance in classifying each label. This can be addressed by adopting balanced accuracy, precision, recall, and F1-score (Iwendi et al., 2020) metrics which provide model performance at each label. Evaluation metrics may help to check the bias of model but oversampling and undersampling techniques are adopted by the research community to overcome data imbalance or long tail data as shown in Table 15.

## 6.6 | Dataset volume

Another challenge in designing an RPM system that uses AI models is the size of the dataset used for its training and predicting purposes. Most machine learning algorithms require large datasets to build a robust model. The size of the dataset matters, as this would hinder the ability of a machine learning model to perform accurately. To analyze hospitalized data or outpatient data, a good model needs to be trained with informative features with a high number of subjects (Ramos et al., 2021). A neural network model could enhance the performance as more data is available (Coppock et al., 2021). Random forests need relatively few training cases to achieve near-peak performance, are computationally cheap to train, and are able to handle large numbers of descriptors well (Teixeira et al., 2016). Data-driven models like logistic regression, SVM, or neural networks have an advantage in model derivation as these models do not require prior knowledge about the relationship between input predictor variables and output target variables. Models like decision trees, random forests, SVM, and Bayesian networks can handle large datasets and integrate background knowledge into the analysis (Awad et al., 2017).

**TABLE 14**  Trends in signal processing

| Challenge | Trends | References |
|---|---|---|
| Signal processing | Fourier transforms | Hou et al. (2017), H. Wang et al. (2016), X. Wang et al. (2017) |
|  | Least mean square (LMS) filter | He et al. (2017), Zhao et al. (2018), Q. Xu et al. (2020) |

**TABLE 15**  Trends in data imbalance

| Challenge | Trends | References |
|---|---|---|
| Imbalanced datasets | OverSampling, SMOTE | Hambali and Gbolagade (2016), Alotaibi and Sasi (2016), Wang, Yao, and Chen (2021), Kumar et al. (2022) |
|  | Undersampling | Choudhary and Shukla (2021), Wang, Yao, and Chen (2021), Kumar et al. (2022) |

**TABLE 16**  Trends in data imbalance

| Challenge | Trends | References |
|---|---|---|
| Feature extraction | Feature engineering feature learning and representation | K. Chen, Zhang, et al. (2021), Kaieski et al. (2020), X. Zhang et al. (2017), Y. Xu et al. (2018) |
|  | Deep learning | Zhong et al. (2016) |

## 6.7 | Feature extraction

Feature extraction is one of the key steps RPM systems perform in analyzing human vital signs and activity recognition (K. Chen, Zhang, et al., 2021). T generate a model to predict, detect or score the patient's health state, the definition of the features must be included (Kaieski et al., 2020). Lack of efficient feature engineering process, feature selection methods, and the heterogeneity of measured patient data are some challenges limiting the effectiveness of machine learning-based predictive models (X. Zhang et al., 2017). Within ICU, patients are monitored continuously by numerous specialized devices at the bedside, which generates high-density multiple data modalities. As a result, the timestamps, order, and frequency of the measurements may be profoundly different from one patient to another. This type of irregularity and heterogeneity in patient data make feature selection even more challenging (Y. Xu et al., 2018). To overcome the lack of efficient feature engineering, feature selection techniques, and the heterogeneity of measured patient data, feature learning or representation learning techniques can be used. Deep learning algorithms such as RNNs, LSTM, CNN, and other algorithms based on neural networks have the capability to learn this type of data structure (Zhong et al., 2016). The feature extraction challenge can be addressed by adopting feature representation techniques. Deep learning can overcome this issue without any additional framework. The research works presenting the trends with corresponding research works are presented in Table 16.

## 7 | FUTURE DIRECTION OF AI ON RPM

The future direction of the research is to extend the scope of AI in RPM applications to enhance healthcare services for both providers and patients. To achieve this, the challenges in adopting AI to RPM as well as AI implementation have to be addressed. As discussed in the previous section, the major challenges in adopting AI are AI or ML explainability, privacy, and uncertainty. Explainability of AI results needs to be improved as it assists healthcare professionals in understanding patients' health status better and helps in decision-making. Explainable AI approaches are working in this direction. Explainability techniques such as SHAP, LIME, DeepLIFT, and so on are being adopted to RPM systems widely. This has to be further improved to breakdown the state-of-the-art deep learning and machine learning results to healthcare practitioners to make informed decisions.

Data privacy and the security of patients' health is another major issue that could be addressed with federated learning. However, there is no strong research evidence in the federated learning concept to confirm that the reverse engineering of the local client model parameters would not lead to a patient's private data. Future works should be focused on strengthening the federated learning framework for data privacy and security. Blockchain technology has proved its capacity in maintaining data privacy with transparency and immutability. However, the implementation of blockchain technology demands high implementation costs and high energy dependence. Future works need to concentrate on blockchain engineering issues.

Uncertainty due to model structure and hyperparameters causes uncertainty in results. UQ technique can play a vital role to reduce uncertainties in the model during optimization and decision-making. The aleatoric and epistemic uncertainty can be addressed through different probabilistic and nonprobabilistic, and inverse uncertainty techniques. Focusing on this challenge would help to improve healthcare professionals' trust in machine learning or deep learning model results.

Data-related challenges are inevitable in AI-based applications. It could be due to IoT devices' signal processing, noise, data imbalance, limited labeled data, and feature extraction. Efficient and clean data is the first and most time-consuming part of AI methodology. The challenges such as label imbalance or long-tail data have to be processed with class balancing discussed in this study. The research community should concentrate on input data related at the most to achieve efficient and effective results.

Reinforcement learning has the potential to mimic human behavior and build social assistive robots for patients in a hospital or at home. Although this AI technique has been used for a while, there has not been much research into applying it to healthcare applications. Taking advantage of the sequential decision-making ability of reinforcement learning, healthcare applications such as dynamic treatment regimens and Just-in-Time-Adaptive-Interventions (JITAIs) can be further enhanced. However, there have been recent incidents of deploying physical robots causing threats to humans (Stein et al., 2022). It is recommended to build virtual robots using reinforcement learning agents to monitor patients and predict unprecedented events.

## 8 | CONCLUSION

Healthcare applications have been widely transformed by technological innovations in information systems and AI. In particular, the last decade has revolutionized monitoring patients' health status by tracking their vital signs and physical activities. Advancements in data transmission and data modeling enabled RPM systems to detect patients' health deterioration in advance, customize patient-centric applications, and learn their behavior patterns adaptively. The transformation of RPM systems using noninvasive information system technologies like telehealth, IoT, cloud, fog, edge, and blockchain are explored in this study. The primary focus of this survey article is to present the role of AI in enhancing RPMs with its ability to learn, predict, and classify patients' behavior and vital signs. Applications of AI in monitoring vital signs, physical activities, chronic diseases, and patient emergencies are investigated. Federated learning facilitates a patient-centric monitoring system to focus on their needs while protecting data privacy. Reinforcement learning enhances RPMs to learn patient behavior patterns in a dynamic environment adaptively. The impact of such advanced AI methodologies on RPM systems is detailed with evidence. Even though AI has the potential to transform RPM services, it has certain challenges like explainability, privacy, and uncertainty. Other than this, data learning challenges include feature extraction, imbalanced labels, data volume, and data processing. In this study, the trends and challenges of AI in RPM are discussed in detail.

This study's limitations are that study is focused on RPM systems with vital signs monitoring and physical activities monitoring but not the electroencephalogram (EEG) monitoring and neurological system-related diseases. Also, this study has not explored all chronic disease monitoring research works. While addressing the limitations and challenges discussed in the study, healthcare applications should adopt advanced technological infrastructures like Cloud/Edge/Fog/Blockchain and AI methods such as Federated Learning and Reinforcement Learning. So far, traditional AI methods such as supervised and unsupervised have demonstrated state-of-the-results. However, this is the right time to transform healthcare for preventive, predictive, and personalized monitoring of patients and provide enhanced assistance to healthcare practitioners.

## AUTHOR CONTRIBUTIONS

**Thanveer Shaik:** Conceptualization (equal); data curation (equal); formal analysis (equal); investigation (equal); methodology (equal); writing – original draft (equal); writing – review and editing (equal). **Xiaohui Tao:** Conceptualization (equal); formal analysis (equal); investigation (equal); methodology (equal); project administration (equal); supervision (equal); writing – original draft (equal); writing – review and editing (equal). **Niall Higgins:** Conceptualization (equal); methodology (equal); supervision (equal); writing – original draft (equal); writing – review and editing (equal). **Lin Li:** Formal analysis (equal); investigation (equal); methodology (equal); writing – review and editing (equal). **Raj Gururajan:** Project administration (equal); supervision (equal); writing – review and editing (equal). **Xujuan Zhou:** Supervision (equal); writing – review and editing (equal). **U. Rajendra Acharya:** Writing – review and editing (equal).

## ACKNOWLEDGMENT
Open access publishing facilitated by University of Southern Queensland, as part of the Wiley - University of Southern Queensland agreement via the Council of Australian University Librarians.

## CONFLICT OF INTEREST
All authors declare there is no conflict of interest in this work.

## DATA AVAILABILITY STATEMENT
Data sharing not applicable to this article as no datasets were generated or analyzed during the current study.

## ORCID
*Thanveer Shaik*  https://orcid.org/0000-0002-9730-665X
*Xiaohui Tao*  https://orcid.org/0000-0002-0020-077X
*Niall Higgins*  https://orcid.org/0000-0002-3260-1711
*Raj Gururajan*  https://orcid.org/0000-0002-5919-0174
*Xujuan Zhou*  https://orcid.org/0000-0002-1736-739X
*U. Rajendra Acharya*  https://orcid.org/0000-0003-2689-8552

## RELATED WIREs ARTICLES
[Internet of Things and data mining: From applications to techniques and systems](#)
[Internet of Things and data analytics: A current review](#)
[Healthcare 4.0: A review of frontiers in digital health](#)